\documentclass[a4paper,12pt]{article}
\usepackage{oagm99}
\usepackage[dvips]{graphicx}

\begin{document}

\title{Web-based Interface in Public Cluster}

\author{{\em Z. Akbar \thanks{~zaenal@teori.fisika.lipi.go.id} and L.T. Handoko \thanks{~handoko@teori.fisika.lipi.go.id}} \\Group for Theoretical and Computational Physics, Research Center for Physics, Indonesian Institute of Sciences \thanks{~http://teori.fisika.lipi.go.id}}

\date{}
\maketitle
\pagestyle{empty}          
\thispagestyle{empty}      

\begin{abstract}%
A web-based interface dedicated for cluster computer which is publicly accessible for free is introduced. The interface plays an important role to enable secure public access, while providing user-friendly computational environment for end-users and easy maintainance for administrators as well. The whole architecture which integrates both aspects of hardware and software is briefly explained. It is argued that the public cluster is globally a unique approach, and could be a new kind of e-learning system especially for parallel programming communities.
\end{abstract}

\section{Introduction}

LIPI Public Cluster (LPC) is a cluster-based computing facility maintained by Lembaga Ilmu Pengetahuan Indonesia - LIPI (the Indonesian Institute of Sciences) \cite{lpc}. Although it is still a small scale cluster in the sense of number of nodes already installed, it has unique characteristics among existing clusters around the globe due to its openness. Here "open" means everyone can access and use it anonymously for free to execute any types of parallel programmings \cite{hakcipta}.

The development of LPC  was initially motivated by real needs for high performance and advanced computing, especially in the research field of basic natural sciences. Even in Indonesia, the needs are growing along with the advances of scientific researches. In the last decades, clustering low specs (and low cost) machines becomes popular to realize an advanced computing system comparable to, or in most cases better than the conventional mainframe-based system with significant cost reduction \cite{Goscinski}.

In general a cluster is designed to perform a single (huge) computational task at certain period. This makes the cluster system is usually exclusive and not at the level of appropriate cost for most potential users, neither young beginners nor small research groups, especially in the developing countries like Indonesia. It is clear that the cluster is in that sense still costly, although there are certain needs to perform such advanced computings. No need to say about educating young generations to be the future users familiar with parallel programmings.  This background motivates us to further develop an open and free cluster environment for public \cite{rict}.

According to its nature LPC is, in contrast with any conventional clusters, designed to accommodate multiple users with their own parallel programmings executed independently at the same period. Therefore an issue on resource allocation is crucial, not only in the sense of allocating hardwares to the appropriate users but also to prevent any interferences among them. In LPC we have deployed a new algorithm to overcome this problem, namely the dependent \cite{iceei} and independent multi-block approaches \cite{icts}. 

Concerning its main objective as a training field to learn parallel programmings, the public cluster should be accessible and user-friendly for all users with various level of knowledges on parallel programming. It also should have enough flexibility regarding various ways of accessing the system in any platforms as well. This can be achieved by deploying web-based interfaces in all aspects. Presently we have resolved some main issues, such as security from anonymous users to prevent any kinds of interference among different tasks running simultaneously on multi blocks \cite{iceei,icts}, algorithm for resource allocation management \cite{icici2} and the real-time monitoring and control over web for both administrators and end-users \cite{icici1}. 

In this paper we first present briefly the concept of LPC including the work flow from the initial registration to the execution of computational works. Thereafter we discuss the main part of this paper, that is the architecture of web-interface in LPC. Finally we conclude with some comments and discussion.

\section{The concept and work flow}

In order to overcome the issues mentioned in the preceeding section, it is clear that we should develope and utilize an integrated web-based interface in LPC. However, concerning real demands in Indonesia presently, we assume that our potential users would be beginners in parallel programmings who are going to use it moreless for educational or self-learning purposes. Although the cluster is also going to be used by some experts and larger research groups to perform more serious advanced computings, we do not expect any anonymous users with heavy computational works. The reason is mainly because of the limited resources, i.e. number of nodes and its specifications. Actually we rather prefer as many as people use our cluster as a training field to learn parallel programmings. In average we plan to provide only $2 \sim 4$ nodes in a relatively short period, namely less than 3 days, for each anonymous user.

These characteristics completely differ with another existing clusters around the world. Because they are usually used by certain people or groups bringing similar types of computational works which are appropriate for the cluster after particular procedures like submitting proposals and any kinds of letter of agreements. In our case, incoming users are completely anonymous and then there is no way in any means to know the type of computational works being executed in the cluster. This fact consequently brings another problems as maintaining job executions owned by different users at same period simultaneously. Under these assumptions and conditions, the public cluster should fulfill very challenging requirements that might be irrelevant in any conventional clusters, that is :
\begin{itemize}
\item Security : \\
This is the main issue we should encounter from the first. The users should have enough priviledges and freedom to optimize their opportunities to learn parallel programming, while their access must be limited at the maximum level for the sake of security of the whole system and another active users at the same period.
\item Flexibility :\\
It is impossible to provide the same level of flexibility for anonymous users as well-defined users with direct accesses through \textsf{ssh}, etc, but we should allow as much as possible the users to execute their codes on LPC. Also there should be a freedom on assigning the number of nodes for each user, since each user could require various number of nodes depending on the computational capacities they actually need. 
\item Stability :\\
Simultaneous executions by different users with various programmes in different blocks of cluster without any interferences among them requires new innovations on cluster management. This problem includes some techniques to incorporate wide range of nodes with different specifications, that is ranging from Intel 486 to the latest Athlon based nodes.
\item Efficiency :\\
Since the cluster is dynamically divided into several blocks with various number of nodes inside according to the users requests, each node should be able to be completely turned on or off partially without any interruptions to another working nodes. 
\end{itemize}
All of these require modifications on some existing tools, and also new developments on alternative softwares and hardwares as well \cite{rict,iceei,icts,icici2,icici1}.

Regarding to the above-mentioned concept, we have defined the work flow in LPC as follows :
\begin{enumerate}
\item A new user should complete an initial registration by providing  personal data, the content of job will be performed and the number of nodes requested for the job.
\item The application is reviewed and verified by the administrator. After approval, the administrator will assign the nodes provided and its usage period. 
\item After reconfirmations by user, proving their agreements with the provided nodes and usage period, the administrator will switch the nodes on that also activate all daemons automatically. 
\item The users should adjust their parallel programmes to fit the provided nodes.
\item The user uploads all necessary programmes and libraries if any. At this stage the programmes can be executed immediately.
\item The administrator and automated system will monitor the usage of all running users.
\item Finished jobs and the results can be downloaded by owners.
\item Once the usage period is over, the nodes are turned off automatically.
\end{enumerate}

Again, these procedures can be done fully and remotely through web. We should mention that the types and number of nodes allocated for a newly assigned block is done automatically utilizing a decision making tool based on the extended genetic algorithm embedded in the web-interface \cite{icici2}. Now we are ready to discuss the architecture of web-interface in LPC. 

\section{The architecture of web-interface}

\begin{figure}[b!]
        \centering \includegraphics[width=14cm]{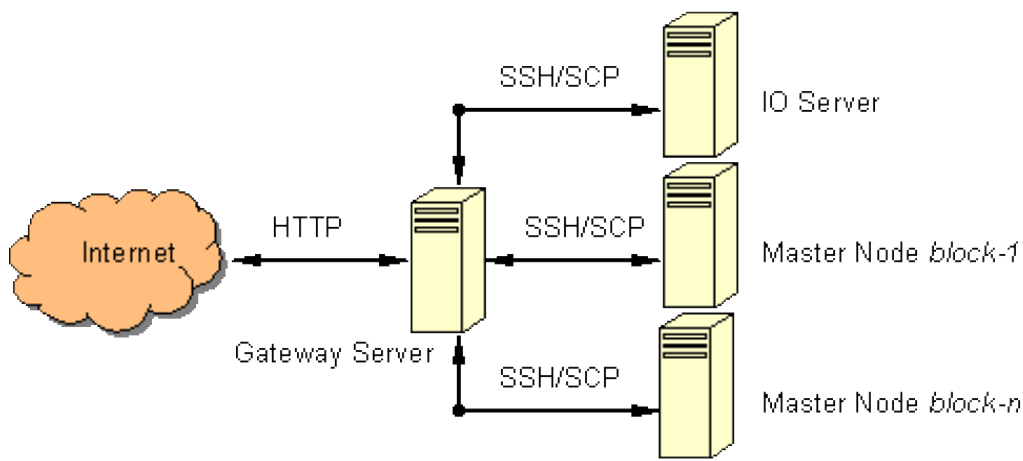}
        \caption{The communication architecture in LPC.}
        \label{fig:komunikasi}
\end{figure}

As mentioned earlier, the web-interface takes over all aspects in LPC, from communication among the nodes till running related tools and softwares to perform computational works. It is also the only way connecting the internal system with the rest of the world. 

In term of communication, there are two kinds of communication in LPC : a) between users and gateway server through HTTP protocol, and b) between gateway server and nodes consisting blocks of cluster through \textsf{ssh} or \textsf{scp}. This architecture is intended to limit direct access to a block of cluster for security reason. On the other hand, this still allows the system to interact with another elements of cluster as IO server and master nodes of each block. As depicted in Fig. \ref{fig:komunikasi}, we put a gateway server as a common entrance for users which serves all related web-based applications. The gateway server further sends and receives commands related to file or parallel programming operations using secure remote login \textsf{ssh} / \textsf{scp}.

\begin{figure}[t]
        \centering \includegraphics[width=14cm]{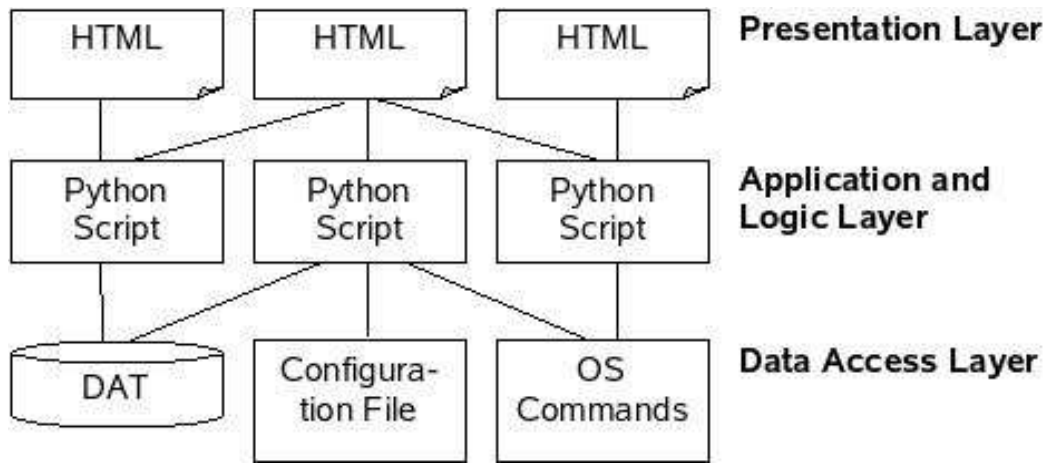}
        \caption{The architecture of three-tier web-based interface in LPC.}
        \label{fig:3layer}
\end{figure}

The web application in LPC should accommodate the needs of any parallel programmings which vary and change along the time. Then we must make it as flexible as possible to guarantee its compatibilities in the future. Also, any future modifications in the web application must not alter another software components. This can be achieved by adopting three-tier software architecture \cite{Foreman}. This method is appropriate for distributed client / server applications which require high performances in terms of flexibility, maintainability, reusability and scalability, while at the same time it simplifies the complexities of users' distributed processes.

This architecture is shown in Fig. \ref{fig:3layer}. As can be seen in the figure, each layer is completely separated which could then simplifies its developments and implementations. Presentation layer behaves as a front interface and interacts directly by receiving inputs from users. The HTML files in this layer could contain either static or dynamic (through application and logic layer) contents. For instance in the case of LPC they are generated dynamically by Python scripts.

The application and logic layer handles all programming logics to serve user needs as uploading the jobs, executing codes and so forth. Further, the data access layer provides the access to data files from the database system, configuration files or generated by any commands in operating system. 

\begin{figure}[t]
        \centering \includegraphics[width=14cm]{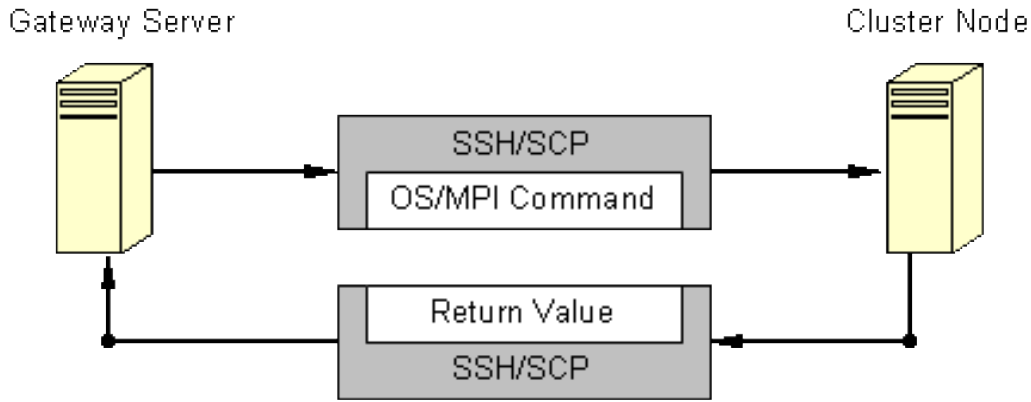}
        \caption{The flow of commands between nodes and gateway server in LPC.}
        \label{fig:command}
\end{figure}

Next, the main task for web-interface in LPC is providing the users an appropriate environment for parallel programming. At time being, there are several parallel programming environments like MPI, PVM and so on \cite{mpi}. Also some modified versions of them as MPICH (version 1 and 2) and LAM-MPI. The web-interface should allow users to choose the preferred one, or to change from one to another. In order to enable dynamic change of environment according to user preferences, we have deployed a mechanism based on the Modules package \cite{Furlani}. This package enables environment switching by changing its path and variables through a single command.

Further issue is how to keep the whole performance of cluster during operation. This can be done by embedding "real-time" control and monitoring system and displaying the results through web. We have developed a dedicated device for this purpose using microcontroller-based hardwares \cite{icici1}. The control system manages the hardware aspect of each node, for instance turning them on or off. The microcontrollers are interfaced by Python and some external programmes in C++. The monitoring system takes the data of external physical observables such casing temperatures and humidities. The physical parameters for each node can be retrieved easily through BIOS. Both control and monitoring systems are integrated to realize an automated hardware control, i.e. shuting down a node if the temperature exceeds the predefined threshold, etc.

Beside these hardware-based management systems, we have also implemented the software-based management system to control all nodes more efficiently. We have deployed a tool set called Cluster Command and Control (C3) \cite{c3} and Ganglia \cite{ganglia}. C3 bundles useful commands to enable automatic processes throughout multiple nodes, while Ganglia is used to monitor the current condition of cluster through web-interface. We have integrated these tools in our web-interface to create a more comprehensive web-based interface matches specific needs in LPC.

Lastly, we should comment on the communication protocol among the nodes in LPC. All commands are sent to nodes through secure shell login \textsf{ssh} / \textsf{scp} passwordlessly using RSA / DSA keys. This handy solution can be used for any commands, either with or without standard outputs. The flow of commands between nodes and user through gateway server is diagrammatically given in Fig. \ref{fig:command}
  
\section{Conclusion}

We have introduced the web-interface in LPC which integrates all aspects of its hardware and software. Although web-based interfaces in clusters are commonly known and actively developed by many groups, the web-interface for public clusters is quite unique and requires more careful developments. Moreover, in any existing conventional clusters, the web-interfaces are moreless complementary tools. While in public clusters like LPC, the web-interface is very crucial and the only solution to make it open for public.

We would like to comment some points regarding the current status of LPC :
\begin{itemize}
\item In some specific cases, the work-flow in LPC can be automated to reduce administration works. For instance, it could be realised for a ``public'' cluster limited to a community with predefined users and types of computational works. However, in our case the present work flow is the simplest and most ``automated'' one regarding the security and its level of openness.
\item Related to this, the main issue in daily operation is to prevent in advanced malicious users who try to submit huge never-ending computational tasks. Actually, this is themain reason we keep current work-flow which is quite powerfull to deal with this potential problem.
\item At time being, the computational results should be downloaded by users. Although it is also possible to make a real-time monitoring for on-going works, we are not going to implement such services in the future. Because it is irrelevant in parallel programmings which could normally take few days computational time. Another reason is such services could exhaust the limited resources of the system, then it is not worthy compared to its importances.
\item The usage time provided for an approved user is determined based on the their request since we do not confirm their code in detail. However, as our basic policy, we provide in most only several days usage time for each anonymous user in one new request.
\item We have implemented our own resource allocation algorithm based on the extended genetic algorithm embedded in the web-interface \cite{icici2}. This is different with the conventional cluster where the resource allocation should be performed during the computational period to achieve the best performance. In the public cluster like LPC, the conventional resource allocation is irrelevant, because it is done only at the first time of node assignment after the approval procedure. Once a block consisting of some resources has been  allocated, it remains unchanged during the usage period. Of course, resource allocation within a block is still relevant, but it uses the existing tools chosen by users. Moreover, doing an appropriate load-balancing within a block used by user is an important aspect of learning parallel programming, so this task should be done by the users themself.
\item The LPC so far enjoys numerous visits and many active users as can be seen in its web \cite{lpc}, ranging from high school to college students. This fact has definitely proven the usability of the system by (mostly) beginners in parallel programmings.
\item The current system is in principle portable to other existing clusters. Especially the concept could be an interesting alternative for some old clusters that might still be usefull for public educational purposes.
\end{itemize}

As future work, we are going to investigate more comprehensively the overall performances, especially in the case of simultaneous heavy computational jobs of multiple users. However, the current performances have been reported partially in our previous works \cite{iceei,icts,icici2}. Also following recent progress in high performance computing, we are in the step of connecting some blocks in LPC into grid computing. 

\section*{Acknowledgement}
This work is financially supported by the Riset Kompetitif LIPI in fiscal year 2007 under Contract no.  11.04/SK/KPPI/II/2007 and the Indonesia Toray Science Foundation Research Grant 2007.

\bibliographystyle{plain}

\begin{thebibliography}{}

\end{thebibliography}


\begin{thebibliography}{99}
\bibitem{lpc}  LIPI Public Cluster, http://www.cluster.lipi.go.id.
\bibitem{hakcipta}  L.T. Handoko : "Public Cluster : mesin paralel terbuka berbasis web", Indonesian Copyright, No. B 268487 (2006).
\bibitem{Goscinski} A. Goscinski, M. Hobbs and J. Silcock : "A Cluster Operating System Supporting Parallel Computing", Cluster Computing, Vol. 4, pp. 145–156 (2001).
\bibitem{rict} Z. Akbar, Slamet, B.I. Ajinagoro, G.J. Ohara, I. Firmansyah, B. Hermanto and L.T. Handoko : "Open and Free Cluster for Public", Proceeding of the International Conference on Rural Information and Communication Technology 2007, Bandung, Indonesia,2007.
\bibitem{iceei} Z. Akbar, Slamet, B.I. Ajinagoro, G.J. Ohara, I. Firmansyah, B. Hermanto and L.T. Handoko : "Public Cluster : parallel machine with multi-block approach", Proceeding of the International Conference on Electrical Engineering and Informatics, Bandung, Indonesia, 2007.
\bibitem{icts} Z. Akbar and L.T. Handoko : "Multi and Independent Block Approach in Public Cluster", Proceeding of the 3rd Information and Communication Technology Seminar, Surabaya, Indonesia, 2007.
\bibitem{icici2} Z. Akbar and L.T. Handoko : "Resource Allocation in Public Cluster with Extended Optimization Algorithm", Proceeding of the International Conference on Instrumentation, Communication, and Information Technology, Bandung, Indonesia, 2007.
\bibitem{icici1} I. Firmansyah, B. Hermanto, Slamet, Hadiyanto and L.T. Handoko : "Real-time control and monitoring system for LIPI Public Cluster", Proceeding of the International Conference on Instrumentation, Communication, and Information Technology, Bandung, Indonesia, 2007.
\bibitem{Foreman} J. Foreman, J. Gross, R. Rosenstein, D. Fisher, K. Brune : "C4 Software Technology Reference Guide - A Prototype", Handbook (CMU/SEI-97-HB-001), Software Engineering Institute, Carnegie Mellon University, January 1997.
\bibitem{mpi} For examples, see : \\
Message Passing Interface Forum : "MPI2: A Message Passing Interface standard". International Journal of High Performance Computing Applications, 12(1–2):1–299, 1998.
\bibitem{Furlani} John L. Furlani `: "Modules: Providing a Flexible User Environment",  Proceedings of the Fifth Large Installation Systems Administration Conference (LISA V), pp. 141-152, San Diego, CA, September 30 - October 3, 1991. http://modules.sourceforge.net.
\bibitem{c3} Ray Flanery, Al Geist, Brian Luethke and Stephen L. Scott : "Cluster Command \& Control (C3) Tool Suite", 3rd Austrian-Hungarian Workship on Distributed and Parallel Systems (DAPSYS 2000) in conjunction with EuroPVM/MPI 2000, Balatonfured, Lake Balaton, Hungary, September 10-13, 2000. http://www.csm.ornl.gov/torc/C3/.
\bibitem{ganglia} Ganglia Monitoring System, http://ganglia.sourceforge.net.
\end{thebibliography}

\end{document}